\documentclass[twocolumn,showpacs,aps,epsfig]{revtex4}

\usepackage{graphicx}
\usepackage{epstopdf}

\usepackage[center]{subfigure}

\begin{document}

\newcommand{\bq}{\begin{equation}}
\newcommand{\eq}{\end{equation}}
\newcommand{\bqn}{\begin{eqnarray}}
\newcommand{\eqn}{\end{eqnarray}}
\newcommand{\nb}{\nonumber}
\newcommand{\lb}{\label}
\newcommand{\noi}{\noindent}
 
\title{Colliding Branes and Formation of Spacetime Singularities}
\author{Andreas Tziolas}
\email{Andreas_Tziolas@baylor.edu}
\affiliation{GCAP-CASPER, Physics Department, Baylor University,
Waco, TX 76798, USA}
\author{Anzhong Wang}
\email{Anzhong_Wang@baylor.edu}
\affiliation{GCAP-CASPER,
Physics Department, Baylor University, Waco, TX 76798, USA} 
 
\begin{abstract}

We construct a class of analytic solutions  with two free parameters 
to the five-dimensional Einstein field equations, which represents
the collision of two timelike 3-branes. We study the local and global 
properties of the spacetime, and find that spacelike  singularities 
generically develop after the collision, due to the mutual focus of
the two branes. Non-singular spacetime can be constructed only in the 
case where both of the two branes violate the energy conditions.

\end{abstract}

\pacs{98.80.Cq, 98.80.-k, 04.40.Nr}
 
\maketitle

\section{Introduction}

 Branes in string/M-Theory are   fundamental constituents \cite{strings}, and of 
particular relevance to cosmology \cite{JST02, branes}. These substances can move
freely in bulk, collide, recoil, reconnect, and whereby form a brane gas in the 
early universe \cite{branegas}, or  create an ekpyrotic/cyclic universe 
\cite{cyc}. Understanding these processes is fundamental to both string/M-Theory
and their applications to cosmology.

Recently, Maeda and his collaborators numerically studied  the collision 
of two branes in a five-dimensional bulk, and found that the formation of a spacelike
singularity after the collision is generic \cite{MT04}. This is a very  
important result, as it implies that a low-energy description of colliding branes 
breaks down at some point, and a complete predictability is lost, without the complete
theory of quantum gravity. Similar conclusions were obtained from the studies of two 
colliding orbifold branes \cite{chen06}. However, lately it was argued that, from the 
point of view of the higher dimensional spacetime where the low effective action was
derived, these singularities are very mild and can be easily regularised \cite{LFT06}.

In this paper, we present a class of analytic solutions to the five-dimensional
Einstein field equations, which represents the collision of two timelike 3-branes 
in a five-dimensional vacuum bulk, and show explicitly that a spacelike singularity 
always develops after the collision due to the mutual focus of the two branes, when
both of them satisfy the energy conditions. If only one of them
satisfies the energy conditions, spacetime singularities  also always exist either 
before  or after the collision. Non-singular spacetimes can  be constructed but
only in the case where both of the two branes violate the energy conditions. 
Specifically, the paper is organized as follows: in the next section we present 
such  solutions, and study their local and global properties, while in Section III  we 
give our main conclusions and remarks.

\section{Colliding  timelike 3-Branes}

\renewcommand{\theequation}{2.\arabic{equation}}
\setcounter{equation}{0}

Let us consider the solutions,
\bq
\lb{2.1}
ds^{2}_{5} = A^{-2/3}(t, y)\left(dt^{2} - dy^{2}\right)
- A^{2/3}(t, y)d\Sigma^{2}_{0},
\eq
where $d\Sigma^{2}_{0} \equiv \left(dx^{2}\right)^{2} + \left(dx^{3}\right)^{2}
+ \left(dx^{4}\right)^{2}$, $\; x^{A} = \left\{t, y, x^{i}\right\},\; (i = 2, 3, 4)$,
 and
\bqn
\lb{2.2}
A(t, y) &=&  a \left(t + b y\right) H\left(t + b y\right)  \nb\\
& &   
+ b \left(t - a y\right) H\left(t - a y\right) \nb\\
& & +   A_{0},
\eqn
with $a, \; b$ and $A_{0}$ being arbitrary constants, and $H(x)$ the
Heavside function, defined as
\bq
\lb{2.2a}
H(x) = \cases{1, & $ x > 0$,\cr
0, & $ x < 0$.\cr}
\eq
Without loss of generality, we assume $a \not= - b$ and $A_{0} > 0$. 
Then, it can be shown that the corresponding spacetime is vacuum, except
 on the hypersurfaces $ t = a y$ and $t = - b y$, where the non-vanishing
components of the Einstein tensor are given by
\bqn
\lb{2.2b}
G_{00} &=& - ab\left(\frac{a\delta(t - ay)}{A} 
                      + \frac{b\delta(t + by)}{A}\right),\nb\\
G_{01} &=&   ab\left(\frac{\delta(t - ay)}{A} 
                      - \frac{\delta(t + by)}{A}\right),\nb\\
G_{11} &=& - \left(\frac{b\delta(t - ay)}{A} 
                      + \frac{a\delta(t + by)}{A}\right),\nb\\
G_{ij} &=& \frac{1}{3}A^{1/3} \delta_{ij} \left(b\left(a^{2} - 1\right)
\delta(t - ay)   \right.\nb\\
& & \left. + a\left(b^{2} - 1\right) \delta(t + by)\right),
\eqn
where $\delta(x)$ denotes the Dirac delta function.
As to be explained below, with the proper choice of the free parameters 
$a$ and $b$,  on each of these two hypersurfaces  the  
spacetime represents  a 3-brane filled with a perfect fluid.

The normal vector to the surfaces $t - a y = 0$
 and $t + b y = 0$ are given, respectively,  by 
\bqn
\lb{2.2c}
n_{A} &\equiv& \frac{\partial (t - ay)}{\partial x^{A}} =
 \delta^{t}_{A} - a \delta^{y}_{A},\nb\\
l_{A} &\equiv& \frac{\partial (t + by)}{\partial x^{A}} =
 \delta^{t}_{A} + b \delta^{y}_{A},
\eqn
for which we find
\bqn
\lb{2.3}
n_{A} n^{A} &=& - A^{2/3}\left(a^{2} - 1\right),\nb\\
l_{A} l^{A} &=& - A^{2/3}\left(b^{2} - 1\right).
\eqn
Thus, in order to have these surfaces be timelike, we must choose
$a$ and $b$ such that 
\bq
\lb{2.3a}
a^{2} > 1, \;\;\;\; 
b^{2} > 1.
\eq
Introducing the timelike vectors $u_{A}$ and $v_{A}$ along 
each of the two 3-branes by
\bqn
\lb{2.4}
u_{A} &=& \frac{1}{A^{1/3}_{u}(t)\left(a^{2} - 1\right)^{1/2}}
\left(a \delta^{t}_{A} - \delta^{y}_{A}\right),\nb\\
v_{A} &=& \frac{1}{A^{1/3}_{v}(t)\left(b^{2} - 1\right)^{1/2}}
       \left(b \delta^{t}_{A} + \delta^{y}_{A}\right),
\eqn
where 
\bqn
\lb{2.5}
A_{u}(t) &\equiv& \left. A(t, y)\right|_{y = t/a} \nb\\
    &=&   \left(a + b \right) t H\left(t + \frac{b}{a} t\right) 
+ A_{0},\nb\\
A_{v} (t) &\equiv& \left. A(t, y)\right|_{y = - t/b} \nb\\
    &=&   \left(a + b \right) t H\left(t + \frac{a}{b} t\right) 
+ A_{0}, 
\eqn
we find $u_{A}n^{A} = 0 = v_{A}l^{A}$. From the five-dimensional Einstein 
field equations, $G_{AB} = \kappa T_{AB}$, we obtain
\bq
\lb{2.6}
T_{AB} =  {A_{u}^{1/3}}T^{(u)}_{AB} \delta(t - a y)
+  {A_{v}^{1/3}}T^{(v)}_{AB} \delta(t + b y),
\eq
where   
\bqn
\lb{2.7}
T^{(u)}_{AB} &=&\rho_{u} u_{A}u_{B} 
+ p_{u}\sum_{i = 2}^{4} {X^{(i, u)}_{A}X^{(i, u)}_{B}}, \nb\\
T^{(v)}_{AB} &=&\rho_{v} v_{A}v_{B} 
+ p_{v}\sum_{i = 2}^{4} {X^{(i,v)}_{A}X^{(i,v)}_{B}},  
\eqn
 $X^{(i, a)}_{A}$   are unit vectors, 
defined as $X^{(i, a)}_{A} \equiv A^{1/3}_{a}\delta^{i}_{A}\; 
(i = 2, 3, 4; \; a = u, v)$, and
\bqn
\lb{2.8}
\rho_{u} &=&   - 3p_{u} = - \frac{b\left(a^{2} - 1\right)}
            {\kappa A^{2/3}_{u}(t)}, \nb\\
\rho_{v} &=&  - 3 p_{v} = - \frac{a\left(b^{2} - 1\right)}
            {\kappa A^{2/3}_{v}(t)}.
\eqn

Therefore, the solutions in the present case represent the collision of
two timelike 3-branes, moving along, respectively, the line $t - a y = 0$ 
and the line $t + b y = 0$.  Each of the two 3-branes supports a perfect fluid. 
They approach each other as $t$ increases, and collide at point $(t, y) 
= (0, 0)$, and then move apart. Depending on the specific values of the 
free parameters $a$ and $b$, we have three distinguishable cases: (a) $\;
a,\; b < -1$; (b) $\; a > 1,\;  b < -1$; and (c) $\; a,\; b > 1$.
The case $a < -1, \; b > 1$ can be obtained from Case (b) by exchanging
the two free parameters. In the following let us consider them separately.

\subsection{$ a < -1, \; b < -1$}

In this subcase, from Eq.(\ref{2.8}) we can see that the perfect fluids on
both of the two branes satisfy all the three energy conditions, weak,
strong, and dominant \cite{HE73}. To study the solutions further, we 
divide the spacetime  into four regions, $\mbox{Region I:} \; t < 0, \; 
{t}/{|b|} < y <  {t}/{|a|}$, $\mbox{Region II:} \; y > 0, \; 
-|a|y < t <  |a|y$, $\mbox{Region III:} \; y < 0, \;  |b|y < t <  - |a|y$,
and $\mbox{Region IV:} \; t > 0, \; -  {t}/{|a|} < y <  {t}/{|b|}$,
as shown in Fig. 1, with the two 3-branes as their boundaries, where we 
denote them, respectively, as, $\Sigma_{u} \equiv \left\{x^{A}: \; t - ay = 
0\right\}$ and $\Sigma_{v} \equiv \left\{x^{A}: \; t + by =0\right\}$.

\begin{figure}
\includegraphics[width=\columnwidth]{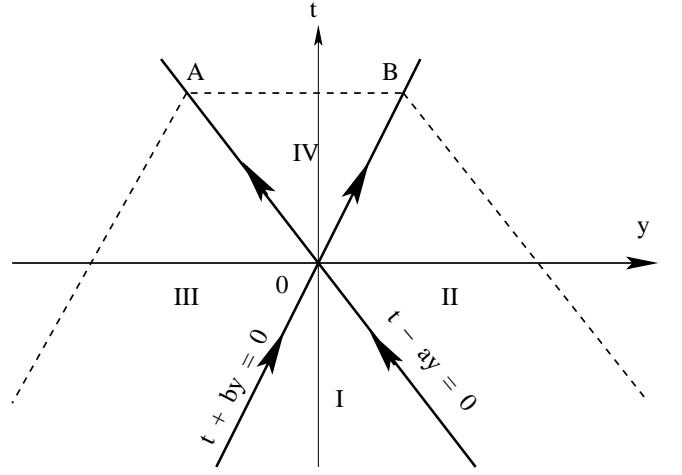}
\caption{The five-dimensional spacetime in the ($t, \; y$)-plane for 
$ a < -1,\; b < -1$.  The two 3-branes  
approach each other from $t = - \infty$ and collide at $(t, y) = (0, 0)$.
Due to their gravitational mutual focus, the spacetime ends up at a spacelike
singularity
on the hypersurface $ A_{0} + (a+b) t = 0$ in Region $IV$, denoted by the
horizontal dashed line. The spacetime is also singular
along the line $A_{0} - |a| (t - |b|y) = 0$ ($A_{0} - |b| (t + |a|y) = 0$) 
in Region $III$ ($II$), which is
parallel to the 3-brane located on the hypersurface $ t+by = 0$ 
($t - ay = 0$).}  
\label{fig1} 
\end{figure} 

Along the hypersurface $\Sigma_{v}$, we find
\bqn
\lb{2.9}
\left. ds^{2}\right|_{t = |b|y} &=& \frac{b^{2} -1}{b^{2}A^{2/3}_{v}(t)}dt^{2}
- A^{2/3}_{v}(t) d\Sigma_{0}^{2} \nb\\
&=& d\tau^{2} - a^{2}_{v}(\tau)d\Sigma_{0}^{2},
\eqn
where
\bqn
\lb{2.10}
A_{v}(t) &=& \cases{A_{0} - \left(|a| + |b|\right)t, & $ t \ge 0$,\cr
A_{0}, & $ t < 0$, \cr}\nb\\
d\tau &=& \frac{\sqrt{b^{2} - 1}}{|b|A^{1/3}_{v}(t)}dt,\nb\\
a_{v}(\tau) &=& \cases{a_{v}^{0}\left(\tau_{0} - \tau\right)^{1/2},
 &  $ t \ge 0$,\cr
A^{1/3}_{0}, & $ t < 0$, \cr}
\eqn
with $\tau_{0} = \tau_{0}\left(a, b, A_{0}\right)$, and $a_{v}^{0} \equiv 
A^{1/3}_{0} \tau^{-1/2}_{0}$. Exchanging the free parameters $a$ and $b$
we can get the corresponding expressions for the brane located on the
hypersurface $t - ay = 0$.  From these expressions and Eq.(\ref{2.8})
we can see that the two 3-branes come from $t = -\infty$ with constant energy
densities and pressures, for which the spacetime on each of the branes is 
Minkowski. After they collide at the point $(t, y) = (0, 0)$, they focus
each other, where $\dot{a}_{v, u} (\tau) < 0$, and finally end up at a 
singularity where ${a}_{v, u} (\tau) = 0$, denoted, respectively, by the
point $A$ and $B$ in Fig. 1.

The spacetime outside the two 3-branes are vacuum, and the function $A(t, y)$
is given by
\bq
\lb{2.11}
A(t, y) =\cases{A_{0} - \left(|a| + |b|\right) t, &   IV,\cr
A_{0} - |a|\left(t - |b|y\right), &   III,\cr
A_{0} - |b|\left(t + |a|y\right), &   II,\cr
A_{0}, &   I.\cr}
\eq
From this expression we can see that the spacetime is Minkowski in Region $I$ and the 
function $A(t, y)$ vanishes on the hypersurfaces $A_{0} - \left(|a| + |b|\right) 
t = 0$ in Region $IV$, $A_{0} - |a|\left(t - |b|y\right) = 0$ in Region $III$,
and $A_{0} - |b|\left(t + |a|y\right) = 0$ in Region $II$, denoted by the dashed
lines in Fig. 1. These hypersurfaces actually represent the spacetime singularities. 
This can be seen clearly from the Kretschmann scalar,
\bqn
\lb{2.12}
I &\equiv& R_{ABCD}R^{ABCD} \nb\\
  &=& \frac{8}{9A^{8/3}} \times \cases{(a+b)^{4}, & IV,\cr
a^{4}\left(b^{2}-1\right)^{2}, & III,\cr
b^{4}\left(a^{2}-1\right)^{2}, & II,\cr
0, & I.\cr}
\eqn

The above analysis shows clearly that, when the matter fields on the two branes satisfy 
the energy conditions, due to their mutual gravitational focus,  a spacelike 
singularity is always formed after the collision. This is similar to the conclusion
obtained by Maeda and his collaborators \cite{MT04}.

\subsection{$ a > 1, \; b < -1$}

In this case,  Eq.(\ref{2.8}) shows that the perfect fluid on
the brane $t = ay$ satisfies all the three energy conditions, while the
one on the brane $t = -b y$ does not. To study these solutions further, 
it is found convenient to consider the two cases $a > |b| > 1$ and 
$|b| > a > 1$ separately.

{\bf Case $2.1)\; a > |b| > 1$:} In this case, the two colliding branes divide
the whole spacetime into the following four regions,
\bqn
\lb{2.13}
 &\mbox{I:}& \; t = \cases{ < ay, & $y < 0$,\cr
                                < |b|y, & $y > 0$,\cr}\nb\\
&\mbox{II:}& \; y < 0, \; ay < t < |b| y,\nb\\
&\mbox{III:}& \; y > 0, \; |b|y < t < a y,\nb\\
&\mbox{IV:}& \; t = \cases{ > ay, & $y > 0$,\cr
                                > |b|y, & $y < 0$,\cr}
\eqn
as shown in Fig. 2. Then, we find that 
\bq
\lb{2.14}
A(t, y) =\cases{A_{0} + \left(a - |b|\right) t, &   IV,\cr
A_{0} + a\left(t - |b|y\right), &   III,\cr
A_{0} - |b|\left(t - a y\right), &   II,\cr
A_{0}, &   I.\cr}
\eq
Clearly, the spacetime is again Minkowski in Region $I$, but the 
function $A(t, y)$ now vanishes only on the hypersurfaces $A_{0} + 
\left(a - |b|\right) t = 0$ in Region $IV$, and $A_{0} - 
|b|\left(t - ay\right) = 0$ in Region $II$, denoted by the dashed 
lines in Fig. 2. Similar to the last case, the Kretschmann scalar
blows up on these surfaces, so they actually represent the spacetime
singularities. As a result, the region $ A_{0}/|b| + ay < t < 
- A_{0}/(a - |b|), \; y < 0$, denoted by   $D$ in Fig. 2, is not
part of the whole spacetime. In Region $III$ we have $A(t, y) > 0$, and no 
any kind of  spacetime singularities appears in this region.

\begin{figure}
\includegraphics[width=\columnwidth]{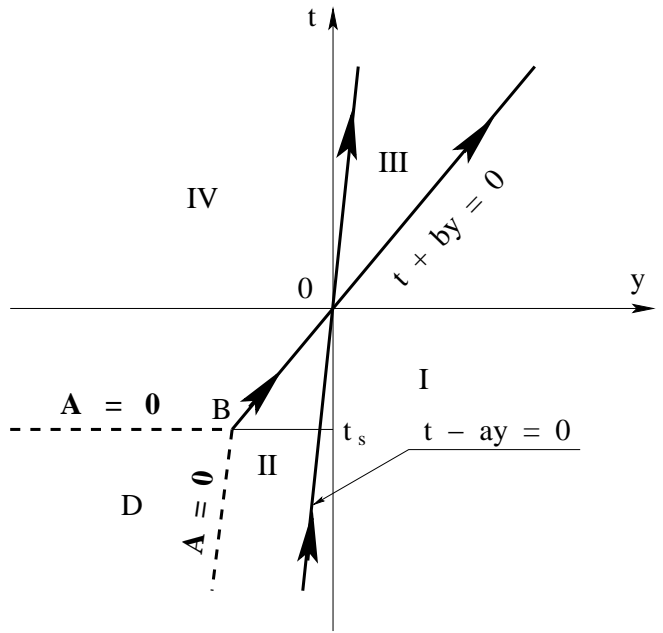}
\caption{The spacetime in the ($t, \; y$)-plane for $ a > |b| > 1,\; b < -1$. 
It is singular along the two half dashed lines, $ t =   - A_{0}/(a  - |b|),
\; y < - A_{0}/[|b|(a  - |b|)]$, and $A_{0} - |b|(t - ay) = 0, \;
t < -  A_{0}/(a  - |b|)$. 
The 3-brane located on the hypersurface $t +by = 0$
starts to expand from the singular point B, $t = -  A_{0}/(a  - |b|)$
and $y = -  A_{0}/[|b|(a  - |b|)]$, until the point $(t, y) =(0, 0)$, where 
it collides  with the other brane moving in along the hypersurface $t-ay = 0$. 
After the collision, it continuously moves forward but with constant energy 
density and pressure, and the spacetime on the brane becomes flat.  
The spacetime on the 3-brane located on the hypersurface $t -ay = 0$ is
flat before the collision, but starts to expand as $a_{u}(\eta) \propto
(\eta + \eta_{0})^{1/2}$ after the collision. This 3-brane is free of any 
kind of spacetime singularities. }  
\label{fig2} 
\end{figure} 

Along the hypersurface $t+by = 0$, the metric takes the same form as that
given by Eq.(\ref{2.9}) but now with  
\bqn
\lb{2.10a}
A_{v}(t) &=& \cases{A_{0}, & $ t \ge 0$, \cr
A_{0} + \left(a - |b|\right)t, & $ t < 0$,\cr}\nb\\
a_{v}(\tau) &=& \cases{A^{1/3}_{0}, & $ t \ge 0$, \cr
a_{v}^{0}\left(\tau + \tau_{s}\right)^{1/2},
 &  $ t < 0$,\cr}
\eqn
where $t = t_{s} \equiv -A_{0}/(a - |b|)$ corresponds to $\tau = \tau_{s}$
and $t = 0$ to $\tau =\tau_{0}$, with $\tau_{0} \equiv 
(b^{2} - )^{1/2}A^{2/3}_{0}/[2|b|(a - |b|)]$, and $a^{0}_{v} = A^{1/3}_{0}
(\tau_{0} + \tau_{s})^{-1/2}$. Thus, in this case the 3-brane located on
the hypersurface $t + by = 0$ starts to expand from the singular point
$\tau = \tau_{s}$ and collides with the other incoming 3-brane at the point
$(t, \; y) = (0, \; 0)$. After the collision, the 3-brane transfers part
of its energy to the one moving along the hypersurface $t - ay = 0$, so that
its energy density and pressure remain constant, and whereby the spacetime on this
3-brane becomes Minkowski.

Along the hypersurface $t - ay = 0$, the metric takes the form
\bqn
\lb{2.9a}
\left. ds^{2}\right|_{t = ay} &=& \frac{a^{2} -1}{a^{2}A^{2/3}_{u}(t)}dt^{2}
- A^{2/3}_{u}(t) d\Sigma_{0}^{2} \nb\\
&=& d\eta^{2} - a^{2}_{u}(\eta)d\Sigma_{0}^{2},
\eqn
where
\bqn
\lb{2.10b}
A_{u}(t) &=& \cases{A_{0} + \left(a - |b|\right)t, & $ t \ge 0$,\cr
A_{0}, & $ t < 0$, \cr}\nb\\
d\eta &=& \sqrt{\frac{a^{2} - 1}{a^{2}A^{2/3}_{u}(t)}}dt,\nb\\
a_{u}(\eta) &=& \cases{a_{u}^{0}\left(\eta + \eta_{0}\right)^{1/2},
 &  $ t \ge 0$,\cr
A^{1/3}_{0}, & $ t < 0$, \cr}
\eqn
where $ t = 0 $ corresponds to $\eta = 0$ and $\eta_{0} \equiv
3(a^{2}-1)^{1/2}A^{2/3}_{0}/[2a(a-|b|)] > 0$. Thus, in the present case
the brane located on the hypersurface $t - ay = 0$ comes from $t = - \infty$
with constant energy density and pressure $\rho_{u} = - 3p_{u}
= |b|(a^{2} -1)/(\kappa A^{2/3}_{0}) >0$, which satisfies all the three
energy conditions. The spacetime on this brane is flat before the collision.
After the collision, the spacetime of the brane starts to expand as 
$(\eta + \eta_{0})^{1/2}$ without   the big-bang type of singularities. 
The expansion rate is the same as  that of a radiation-dominated  universe 
in Einstein's theory of 4D gravity, where $a(\eta)  \propto \eta^{1/2}$. But 
its energy density and pressure now decreases as $\rho_{u} = - 3p_{u} 
\propto (\eta   + \eta_{0})^{-1}$, in contrast to $\rho = 3 p   \propto 
\eta^{-2}$ in Einstein's  4D gravity \cite{HE73}.

{\bf Case $2.2)\;  |b| > a > 1$:} In this case, the two colliding branes divide
the whole spacetime into the four regions,
\bqn
\lb{2.13a}
 &\mbox{I:}& \; t = \cases{ < |b|y, & $y < 0$,\cr
                            < ay, & $y > 0$,\cr}\nb\\
&\mbox{II:}& \; y < 0, \; |b|y < t < a y,\nb\\
&\mbox{III:}& \; y > 0, \; a y < t < |b|y,\nb\\
&\mbox{IV:}& \; t = \cases{ > |b|y, & $y > 0$,\cr
                            > a y, & $y < 0$,\cr}
\eqn
as shown in Fig. 3. 

\begin{figure}
\includegraphics[width=\columnwidth]{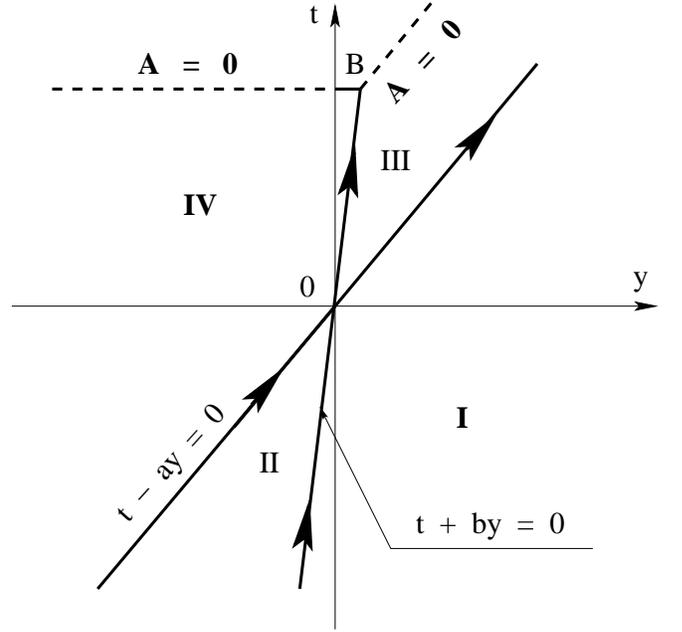}
\caption{The spacetime in the ($t, \; y$)-plane for $ |b| > a > 1,\; b < -1$. 
It is singular along the two half dashed lines where $A = 0$.  The spacetime
of the 3-brane along $t+by = 0$ is flat before the collision, but collapses
to form a spacetime singularity at the point $B$. The spacetime
of the 3-brane along $t- ay = 0$ is contracting from $t = - \infty$ before the 
collision, but becomes flat after the collision. At the colliding point $(t, y)
= (0, 0)$ no any kind of spacetime singularities exists.
 }  
\label{fig3} 
\end{figure} 

Following a similar analysis as we did in the last
subcase one can show that the spacetime now is singular on the half
lines  $t = A_{0}/\left(|b| - a\right),\; y < A_{0}/[|b|\left(|b| - a\right)]$
in Region $IV$, and $ t = A_{0}/|b| + ay > A_{0}/\left(|b| - a\right)$ in 
Region $III$, denoted by  the dashed lines in Fig. 3. 

Along the hypersurface $t - ay = 0$, the metric takes the   form of 
Eq.(\ref{2.9a}) but now with 
\bqn
\lb{2.10c}
A_{u}(t) &=& \cases{A_{0}, & $ t \ge 0$,\cr
A_{0} - \left(|b|- a\right)t, & $ t < 0$,\cr}\nb\\
a_{u}(\eta) &=& \cases{A^{1/3}_{0}, &  $ t \ge 0$,\cr
a_{u}^{0}\left(\eta_{0} - \eta\right)^{1/2},  & $ t < 0$, \cr}
\eqn
where $ t \le 0 $ corresponds to $\eta \le 0$ with $\eta_{0} \equiv
3(a^{2}-1)^{1/2}A^{2/3}_{0}/[2a(|b| -a)] > 0$. Thus, in the present case
the brane located on the hypersurface $t - ay = 0$ comes from $t = - \infty$
with energy density and pressure $\rho_{u} = - 3p_{u}
\propto   \left(\eta_{0} - \eta\right)^{-1}$, which satisfies all the three
energy conditions. The spacetime on this brane is non-flat before the collision
and becomes flat after the collision.

Along the line $t+by = 0$, the metric takes the same form as that
given by Eq.(\ref{2.9}) but now with  
\bqn
\lb{2.10d}
A_{v}(t) &=& \cases{A_{0} - \left(|b| - a\right)t, & $ t \ge 0$, \cr
A_{0}, & $ t < 0$,\cr}\nb\\
a_{v}(\tau) &=& \cases{a_{v}^{0}\left(\tau_{s} - \tau \right)^{1/2}, 
  & $ t \ge 0$, \cr
A^{1/3}_{0}, &  $ t < 0$,\cr}
\eqn
where $t = t_{s} \equiv  A_{0}/(|b| -a)$ corresponds to $\tau = \tau_{s}$
and $t = 0$ to $\tau =\tau_{0}$, with $\tau_{0} \equiv 
(b^{2} - )^{1/2}A^{2/3}_{0}/[2|b|(|b|-a)]$. Thus, in this case the 3-brane 
located on the hypersurface $t + by = 0$ moves in from $t = - \infty$ and
has constant energy density and pressure before the collision. After the
collision, it collapses to a singularity at  
$\tau = \tau_{s}$.

\subsection{$ a > 1, \; b > 1$}

In this subcase, from Eq.(\ref{2.8}) we can see that both of the two branes 
violate all  the three energy conditions \cite{HE73}. Dividing
the spacetime into the following four regions,
\bqn
\lb{2.13b}
&\mbox{I:}& \; t < 0, \; \frac{t}{a} < y < - \frac{t}{b},\nb\\
&\mbox{II:}& \; y > 0, \; -by < t < a y,\nb\\
&\mbox{III:}& \; y < 0, \; ay < t < -b y,\nb\\
&\mbox{IV:}& \; t > 0, \; - \frac{t}{b} < y <  \frac{t}{a},
\eqn
as shown in Fig. 4, we find that 
\bqn
\lb{2.14b}
A(t, y) &=& \cases{A_{0} + \left(a + b\right) t, &   IV,\cr
A_{0} + b\left(t - a y\right), &   III,\cr
A_{0} + a\left(t +b y\right), &   II,\cr
A_{0}, &   I,\cr}\nb\\
A_{u}(t) &=& \cases{ A_{0} + (a+b)t, & $t \ge 0$,\cr
                     A_{0}, & $t < 0$,\cr}\nb\\
A_{v}(t) &=& \cases{ A_{0} + (a+b)t, & $t \ge 0$,\cr
                     A_{0}, & $t < 0$,\cr}                    
\eqn
which are non-zero in the whole spacetime. Thus, in the present case the spacetime
is free of any kind of spacetime singularities, and flat in Region $I$.  Before 
the collision the two branes
move in from $t = -\infty$ all with constant energy density and pressure.   
After the collision, their energy densities and pressures all decrease like 
$\tau^{-1}$, while the spacetime on these two branes is expanding like $a(\tau) 
\propto \tau^{1/2}$,
where $\tau$ is the proper time on each of the two branes, and $a(\tau)$ their
expansion factor. 

\begin{figure}
\includegraphics[width=\columnwidth]{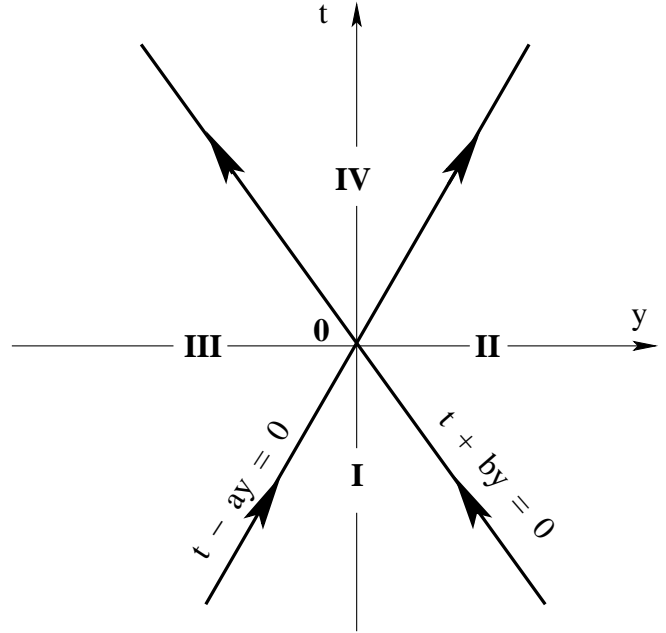}
\caption{The spacetime in the ($t, \; y$)-plane for $ a >  1,\; b > 1$. 
It is free of any kind of spacetime singularities in the whole spacetime,
including the two hypersurfaces of the 3-branes.  The two 3-branes all come
from $t = -\infty$ with constant energy density and pressure. They remain
so  until the moment right before collision. After the 
collision, the spacetime on each of the 3-branes is expanding like
$a(\tau) \propto \tau^{1/2}$, while their  energy densities and pressures
decrease like $\rho = - 3p \propto \tau^{-1}$. }  
\label{fig4} 
\end{figure} 

\section{Conclusions}

In this paper, we have studied the collision of branes and the formation of spacetime
singularities. We have constructed a class of analytic solutions 
to the five-dimensional Einstein field equations, which represents such a collision,
and found that when both of the two 3-branes satisfy the energy conditions, a spacelike
singularity is always developed after the collision, due to their mutual gravitational 
focus. This is consistent with the results obtained by Maeda and his collaborators 
\cite{MT04}. When only one of the two branes satisfies the energy conditions, 
the other brane either starts to expands from a singular point [cf. Fig. 2], or comes 
from $t = - \infty$ and then focuses to a singular point after the collision [cf. Fig. 3]. 
However, if both of the two colliding 3-branes violate the weak energy condition, 
no spacetime singularities exist at all in the whole spacetime. Before the collision, the
two branes approach each other in a flat background with constant energy densities 
and pressures. After they collide at $(t, y) = (0, 0)$, they start to expand as
$a(\tau) \propto \tau^{1/2}$, where $a(\tau)$ denotes their expansion factor, and 
$\tau$ their proper time. As the branes are  expanding, their energy densities 
and pressures decrease as $\rho, \; p \propto \tau^{-1}$, in contrast to that of
$\rho, \; p \propto \tau^{-2}$ in the four-dimensional FRW model.  

As argued in \cite{LFT06}, these singularities may become very mild when the 
five-dimensional spacetime is left to higher dimensional spacetimes, ten dimensions in 
string theory and eleven in M-Theory, a question that is under our current investigation.
 
\section*{ACKNOWLEDGMENTS} 

The financial assistance from the vice provost office for research at Baylor University is 
kindly acknowledged.

\end{document}